# Face Synthesis with Landmark Points from Generative Adversarial Networks and Inverse Latent Space Mapping

Shabab Bazrafkan, *Student, IEEE*, Hossein Javidnia, *Student, IEEE* and Peter Corcoran, *Fellow, IEEE*

*Abstract*— Facial landmarks refer to the localization of fundamental facial points on face images. There have been a tremendous amount of attempts to detect these points from facial images however, there has never been an attempt to synthesize a random face and generate its corresponding facial landmarks. This paper presents a framework for augmenting a dataset in a latent $\mathcal{Z}$-space and applied to the regression problem of generating a corresponding set of landmarks from a 2D facial dataset. The BEGAN framework has been used to train a face generator from CelebA database. The inverse of the generator is implemented using an Adam optimizer to generate the latent vector corresponding to each facial image, and a lightweight deep neural network is trained to map latent $\mathcal{Z}$-space vectors to the landmark space. Initial results are promising and provide a generic methodology to augment annotated image datasets with additional intermediate samples.

*Index Terms*— Generative Adversarial Networks, Database Augmentation, Facial Landmark

## I. INTRODUCTION

Facial landmark detection is a key step in a variety of applications such as emotion recognition [1], expression analysis [2], and face recognition [3]. A significant amount of effort has been invested during the past decade to automate the estimation of facial landmarks. Nevertheless, this task remains an open challenge for datasets involving unconstrained facial poses, complex expressions, and variable lighting conditions. Recently, several researchers have successfully employed deep learning and convolutional neural network (CNN) methods due to the powerful ability of such networks to handle nonlinear data for facial landmark detection [4]–[8].

One of the principle challenges in deep learning research projects is a lack of representative data. Neural networks learn in a progressive manner, and training an accurate model requires a significant set of annotated data. Improving on existing models requires more data, but due to the widespread usage of existing public datasets such as Multi-PIE [9], AFW [10], HELEN [11], LFPW [12] these dataset are no longer useful to improve facial models due to several limitations: (i) these datasets contain a limited number of subjects; (ii) images are acquired with the controlled facial pose and under consistent illumination conditions; (iii) the accuracy of ground truth landmarks is limited due to margins of error in the manual annotation process; also the number of landmarks and their locations is not consistent across datasets. Thus while there are many annotated public facial datasets available to researchers they cannot be used together to improve the quality and accuracy of facial modeling.

One approach that has proved effective in the data preparation phase of the deep learning process is data augmentation [13]. This requires adjusting the characteristics of the original training data, such as contrast, exposure level, or in-plane rotation. However, augmentation of existing facial datasets cannot increase the number of participants, or generate additional 3D facial poses, or 3D lighting conditions and it is a highly time-consuming process. In [13] the authors introduce a smart augmentation technique wherein two or more samples are mixed in an abstract space, and a new sample from the same class is generated. This method is only applicable to classification problems. Expanding a complete facial database in such a transform space, including their corresponding landmarks has not been considered to date in the literature. The goal of this current work is to explore the potential for generating structured random faces that incorporate repeatable facial landmarks using an approach based on Generative Adversarial Networks (GAN).

The most relevant research work is that of Gender Preserving Generative Adversarial Network (GP-GAN) [14] where adversarial networks are exploited to synthesize faces from the landmarks. The generator sub-network in GP-GAN is based on UNet [15] and DenseNet [16] architectures while the discriminator sub-network is based on [17]. Note that the network is using a new gender-preserving loss in parallel with the perceptual loss.

Another relevant study, Age-cGAN [18], employs a GAN to generate random faces and corresponding facial metadata. The focus of Age-cGAN is on identity-preserving face aging where the person's facial attributes are altered to age his/her face while the identity is preserved. The generator and discriminator sub-networks of Age-cGAN have the same architecture as [19]. This network can be used to synthesise augmented facial datasets incorporating aging of subjects.

In the present paper, we propose a GAN framework to generate randomized faces along with the corresponding facial landmarks. To the best of our knowledge, the framework proposed herein is the first description of a GAN that enables researchers to synthesize a fully-annotated random facial

The authors would like to thank Joseph Lemley for his helpful comments.

The research work presented here was funded under the Strategic Partnership Program of Science Foundation Ireland (SFI) and co-funded by SFI and FotoNation Ltd. Project ID: 13/SPP/I2868 on "*Next Generation Imaging for Smartphone and Embedded Platforms*".

S. B. is with the Department of Electronic Engineering, College of Engineering, National University of Ireland Galway, University Road, Galway, Ireland. (e-mail: s.bazrafkan1@nuigalway.ie).

H. J. is with the Department of Electronic Engineering, College of Engineering, National University of Ireland Galway, University Road, Galway, Ireland. (e-mail: h.javidnia1@nuigalway.ie).

P. C. is with the Department of Electronic Engineering, College of Engineering, National University of Ireland Galway, University Road, Galway, Ireland. (e-mail: peter.corcoran@nuigalway.ie).

dataset incorporating a wide range of variations in pose, illumination and facial appearance.

## II. GENERATIVE ADVERSARIAL NETWORK

In this work, the Boundary Equilibrium Generative Adversarial Network (BEGAN), presented in [20] is implemented and trained on a subset of CelebA dataset explained in section IV. In this approach, the discriminator network is an auto-encoder, and the generator architecture is the same as the encoder part of the discriminator. The encoder and decoder parts are shown in Fig 1 and Fig 2 respectively.

In the encoder, all kernels are 3×3, and ELU [21] nonlinearity is used in all layers apart from the red layers where the kernel size is 1×1, and no nonlinearities are employed. No non-linearity is applied to the fully connected layers. In the decoder network, all convolutional layers have 64 channels while in the encoder, the number of the channels is gradually increased to 128, 192, and 256 after each pooling layer.

Suppose that $x$ is the real data coming from the database, $z$ is a sample from the uniformly distributed random space $\mathcal{Z}$, $\mathcal{D}$ is the auto-encoder function with the loss defined by:

$$\mathcal{L}(v) = |v - \mathcal{D}(v)|^2 \quad (1)$$

where $v$ is the input to the auto-encoder. The objectives for BEGAN given by [20] are:

$$\begin{cases} \mathcal{L}_D = \mathcal{L}(x) - k_t.\mathcal{L}\big(G(z_D)\big) & \text{for } \theta_D \\ \mathcal{L}_G = \mathcal{L}\big(G(z_G)\big) & \text{for } \theta_G \\ k_{t+1} = k_t + \lambda_k \left(\gamma \mathcal{L}(x) - \mathcal{L}\big(G(z_G)\big)\right) & \text{for each training step } t \end{cases} \quad (2)$$

where $\mathcal{L}_D$ is the discriminator loss, $\mathcal{L}_G$ is the generator loss, $G(v)$ is the output of the generator for input vector $v$, $\gamma$ is the equilibrium hyper parameter set to 0.5 in this work and $\lambda_k$ is the learning rate for $k$. The ADAM optimizer is used with learning rate, $\beta_1$, and $\beta_2$ equal to 0.0001, 0.5, and 0.999, respectively. BEGAN was trained with the Lasagne library on the top of Theano library in Python. The results for BEGAN are shown in Fig. 3.

## III. INVERSE OF GENERATOR

Since the generator learned the distribution of the database, as it is shown in [20], it can map an $N_z$ dimension random vector into an interpolation point of the database distribution.

The reverse applies as well. Having an image from the dataset, one can estimate the sample in the $\mathcal{Z}$-space ($z_r$), which can produce the image. The method is given in [20] wherein the sample from the $\mathcal{Z}$-space is approximated by optimizing the error function:

$$err = |x_s - G(z_r)| \quad (3)$$

wherein $x_s$ is the sample image and $G$ is the generator function. The ADAM optimizer is used to solve the problem with learning rate, $\beta_1$, and $\beta_2$ equal to 0.1, 0.9, and 0.999, respectively. To check this method, one example is given in Fig. 4.

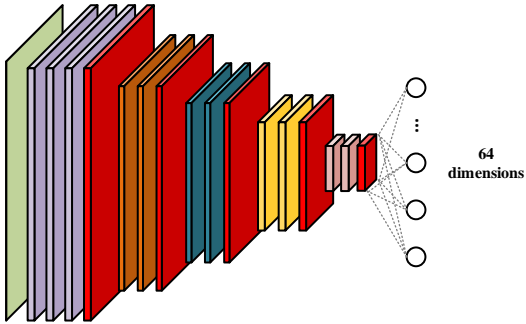

Fig 1. Encoder architecture for BEGAN

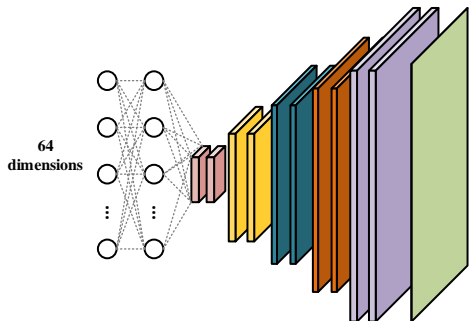

Fig 2. Decoder architecture for BEGAN

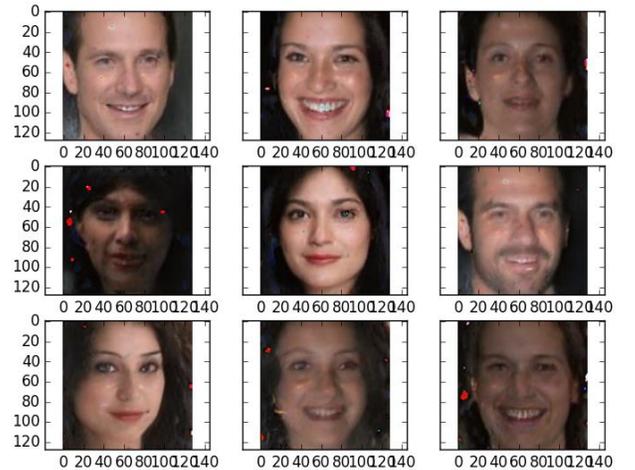

Fig. 3. Generating a random set of images from using BEGAN framework

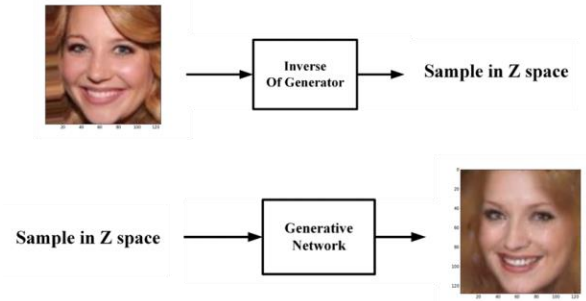

Fig. 4. The inverse of the generator is able to produce an accurate estimate of z vector correspond to each image in the database.

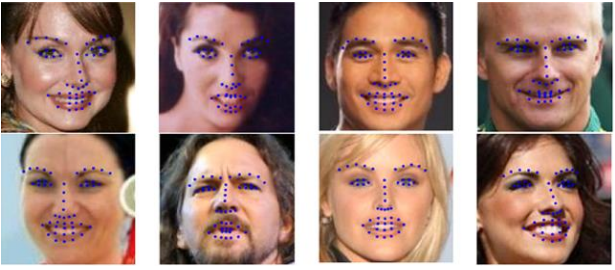

Fig. 5. Facial landmark detection from the discriminative deformable model [24]

## IV. DATABASE

The CelebA dataset [22] consisting of 202,599 original images with 40 unique attributes is used for training our GAN framework. The OpenCV frontal face cascade classifier [23] is used to detect facial regions which are cropped and resized to 128×128 pixels. Initial landmark detection is performed using the method presented in [24] due to its ability to be effective on unconstrained faces. Authors in [24] have augmented the original cascade regression framework of [25] by proposing an incremental algorithm for cascade regression learning. This method personalizes the Supervised Descent Method (SDM) [26] for facial point localization, initializing the SDM offline on a large database of faces, and using newly tracked faces to update it incrementally. The detector in [24] uses a discriminative 3D facial deformable model fitted to the 2D image. The detector was trained on the 300W dataset [27]. It estimates a set of 49 landmarks defined on the contours of eyebrows, eyes, mouth and the nose as shown in Fig. 5.

## V. PROPOSED METHOD

Inverse transform sampling theorem declares that by knowing the probability distribution of a random variable $x_r$, there is a transformation, mapping a uniformly distributed random variable $z_r$ into the space of $x_r$. Since the generator network in the GAN accepts a uniform random variable and transforms it to the image space, this network is learning an approximation of the data distribution. This provides justification for the ability of the neural network to learn the distribution of the landmarks for a given generator network. Our results demonstrate that it is indeed possible to train a network that learns the distribution of landmarks and can map the same uniform random space into the landmark space.

Training a new GAN that learns only the landmark distribution is not going to solve this problem, because this landmark generator is mapping a new uniformly random space $\mathcal{Z}_l$ (which is different from $\mathcal{Z}$-space) into the landmark space. What is required is to map the $\mathcal{Z}$-space used in the BEGAN framework into the landmark space. To achieve this mapping, the following approach is proposed.

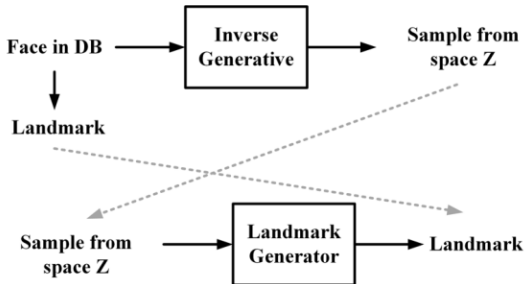

Fig. 6. Proposed method; the landmark generator maps $\mathcal{Z}$-space into a "learned" landmark space

TABLE I
THE ARCHITECTURE OF THE L-GEN. IT IS A SMALL FULLY CONNECTED DEEP NEURAL NETWORK

| Layer name | Layer kind | Number of nodes | Activation |
|---|---|---|---|
| Input Layer | Input | 64 | --- |
| First Hidden | Fully connected | 128 | RELU |
| Second Hidden | Fully connected | 128 | RELU |
| Third Hidden | Fully connected | 128 | RELU |
| Output Layer | Fully connected | 98 | Sigmoid |

Using the techniques described in section IV the landmark set for each face sample of the CelebA database is calculated, and the latent vector in $\mathcal{Z}$-space corresponding to each image is estimated using the inverse generator methodology described in section III. Having these two sets of data (landmarks and $z_r$ for each sample), a lightweight deep neural network can be trained to map each sample in $\mathcal{Z}$-space into a corresponding location in landmark-space. Thus the uniformly distributed space ($\mathcal{Z}$), used in BEGAN to generate faces, is mapped to the landmark space as illustrated in Fig. 6.

THE ARCHITECTURE OF THE LANDMARK GENERATOR (L-GEN) NETWORK TRAINED TO APPROXIMATE THE LANDMARKS IS SHOWN IN

Table I. This network accepts 64 dimensional samples from the $\mathcal{Z}$-space and the output is a set of 98 dimensions corresponding to 49 2D landmark points. The loss function for this network is the Mean Square Error given by:

$$loss = \frac{1}{B_N \times 98} \sum_{k=1}^{B_N} \sum_{i=1}^{98} (o_i - t_i)^2 \quad (4)$$

wherein $o_i$ is the $i$'th output of the output layer, $t_i$ is the $i$'th target value, and $B_N$ is the batch size which set to 16. The Adam optimizer is used to train the network with learning rate, $\beta_1$, and $\beta_2$ equal to 0.0003, 0.9, and 0.999 respectively. The training was done for 1000 epochs using all data. No validation and test set were used in the method.

Adding validation and test sets made the results less accurate since after reducing the number of samples in the training set the network was blind to some samples and could not reconstruct the landmark distribution accurately. To the best of our knowledge, this is the first attempt to generate samples and their corresponding landmarks at the same time.

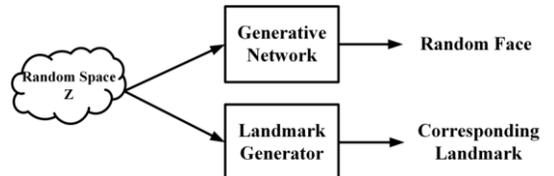

Fig. 7. Feedforward of the proposed method. The uniformly distributed random vector is fed to the Generative network given by BEGAN, and the L-Gen from the proposed method is producing the landmark positions for the generated face image

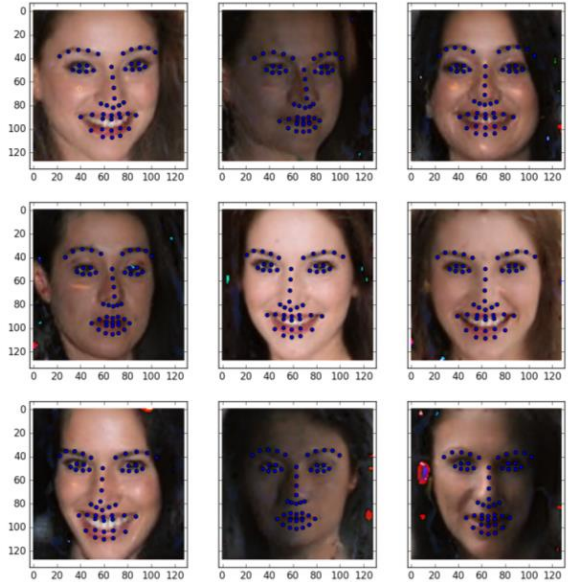

Fig. 8. Random BEGAN faces and generated landmarks.

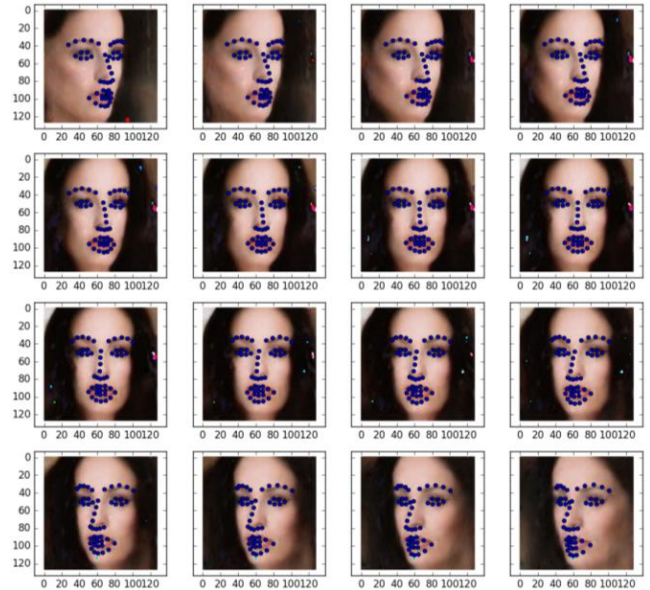

Fig. 9. Interpolating in the Z-space between a face and its mirror image indicates strong continuity for both of these generators

## VI. RESULTS

In the feedforward step shown in Fig. 7, a uniformly distributed random vector is fed concurrently into the GAN (BEGAN) and the L-Gen. The GAN generates a random interpolated face, the Landmark Generator provides a 2D set of landmarks corresponding to the generated face.

Fig. 8 shows example results. Initial results show the L-Gen has learned to map a set of landmark points from the same $\mathcal{Z}$-space as the facial generator, with good generalization across varying pose & illumination conditions.

In [20] the authors investigate the continuity of the face distribution given by the BEGAN face generator by feeding a random set of numbers interpolating two samples in $\mathcal{Z}$-space and observe the gradual changes from one face to another. To investigate the continuity of the landmark estimator distribution, the same approach is used. The $z_r$ vector for a given image and its mirror image is estimated using the inverse generator method, and 14 interpolation points between the two $\mathcal{Z}$-space samples are fed into the face generator and L-Gen networks.

One example result from this experiment is shown in Fig. 9. This figure shows that the landmark distribution estimated by the L-Gen network is smooth in the $\mathcal{Z}$ space.

## VII. CONCLUSION

Data augmentation is a crucial step for modern machine learning frameworks including deep learning approaches. New deep neural networks need a large number of samples to be trained in order to avoid overfitting. The augmentation process introduces a certain amount of uncertainty into the database which helps the network avoid overfitting and generalize the results.

To the best of our knowledge, augmentations presented for regression problems are all applied in the image space. These operations include flipping, rotating, manipulating the contrast and illumination of the image and applying distortions to the image. In this article, a framework for augmenting the database in a latent space is presented and is applied to a regression problem (generating facial landmarks). The BEGAN framework has been used to train a face generator from the CelebA database. The inverse of the generator was implemented using an Adam optimizer to generate the latent vector corresponding to each image, and a small deep neural network was deployed to map the latent vector to the landmark space.

Our observations show that the mapping for both the face generator and L-Gen is continuous and smooth in the latent space. This gives the opportunity to generate a large number of facial samples and their corresponding landmarks, thus expanding the database by introducing more variations through the generation of multiple intermediate samples. The primary goal of this work is to present a framework to expand a database in latent space for regression problems. Our initial studies show very promising outcomes.

Future work will include using the hand-crafted landmarks from the original database, which gives higher accuracy results; additionally, an ongoing empirical study based on the augmented dataset will confirm the efficacy of this technique to generalize from the original CelebA dataset.

## REFERENCES

[1] D. Cristinacce and T. F. Cootes, "Feature Detection and Tracking with Constrained Local Models.," in *BMVC*, 2006, vol. 1, no. 2, p. 3.

[2] S. Taheri, P. Turaga, and R. Chellappa, "Towards view-invariant expression analysis using analytic shape manifolds," in *Face and Gesture 2011*, 2011, pp. 306–313.

[3] V. Bettadapura, "Face expression recognition and analysis: the state of the art," *arXiv Prepr. arXiv1203.6722*, 2012.

[4] Y. Wu, T. Hassner, K. Kim, G. Medioni, and P. Natarajan, "Facial landmark detection with tweaked convolutional neural networks," *arXiv Prepr. arXiv1511.04031*, 2015.

[5] Z. He, J. Zhang, M. Kan, S. Shan, and X. Chen, "Robust FEC-CNN: A High Accuracy Facial Landmark Detection System," in *2017 IEEE Conference on Computer Vision and Pattern Recognition Workshops (CVPRW)*, 2017, pp. 2044–2050.

[6] Z. He, M. Kan, J. Zhang, X. Chen, and S. Shan, "A Fully End-to-

End Cascaded CNN for Facial Landmark Detection," in *2017 12th IEEE International Conference on Automatic Face & Gesture Recognition (FG 2017)*, 2017, pp. 200–207.

[7] A. Jourabloo and X. Liu, "Large-Pose Face Alignment via CNN-Based Dense 3D Model Fitting," in *2016 IEEE Conference on Computer Vision and Pattern Recognition (CVPR)*, 2016, pp. 4188–4196.

[8] S. Xiao, J. Feng, J. Xing, H. Lai, S. Yan, and A. Kassim, "Robust Facial Landmark Detection via Recurrent Attentive-Refinement Networks," in *Computer Vision -- ECCV 2016: 14th European Conference, Amsterdam, The Netherlands, October 11--14, 2016, Proceedings, Part I*, B. Leibe, J. Matas, N. Sebe, and M. Welling, Eds. Cham: Springer International Publishing, 2016, pp. 57–72.

[9] R. Gross, I. Matthews, J. Cohn, T. Kanade, and S. Baker, "Multi-PIE," *Image Vis. Comput.*, vol. 28, no. 5, pp. 807–813, 2010.

[10] X. Zhu and D. Ramanan, "Face detection, pose estimation, and landmark localization in the wild," in *Proceedings of the IEEE Computer Society Conference on Computer Vision and Pattern Recognition*, 2012, pp. 2879–2886.

[11] V. Le, J. Brandt, Z. Lin, L. Bourdev, and T. S. Huang, "Interactive Facial Feature Localization," in *Computer Vision -- ECCV 2012: 12th European Conference on Computer Vision, Florence, Italy, October 7-13, 2012, Proceedings, Part III*, A. Fitzgibbon, S. Lazebnik, P. Perona, Y. Sato, and C. Schmid, Eds. Berlin, Heidelberg: Springer Berlin Heidelberg, 2012, pp. 679–692.

[12] P. N. Belhumeur, D. W. Jacobs, D. J. Kriegman, and N. Kumar, "Localizing parts of faces using a consensus of exemplars," in *CVPR 2011*, 2011, pp. 545–552.

[13] J. Lemley, S. Bazrafkan, and P. Corcoran, "Smart Augmentation Learning an Optimal Data Augmentation Strategy," *IEEE Access*, vol. 5. pp. 5858–5869, 2017.

[14] X. Di, V. A. Sindagi, and V. M. Patel, "GP-GAN: Gender Preserving GAN for Synthesizing Faces from Landmarks," *arXiv Prepr. arXiv1710.00962*, 2017.

[15] O. Ronneberger, P. Fischer, and T. Brox, "U-Net: Convolutional Networks for Biomedical Image Segmentation," in *Medical Image Computing and Computer-Assisted Intervention -- MICCAI 2015: 18th International Conference, Munich, Germany, October 5-9, 2015, Proceedings, Part III*, N. Navab, J. Hornegger, W. M. Wells, and A. F. Frangi, Eds. Cham: Springer International Publishing, 2015, pp. 234–241.

[16] G. Huang, Z. Liu, L. v. d. Maaten, and K. Q. Weinberger, "Densely Connected Convolutional Networks," in *2017 IEEE Conference on Computer Vision and Pattern Recognition (CVPR)*, 2017, pp. 2261–2269.

[17] P. Isola, J.-Y. Zhu, T. Zhou, and A. A. Efros, "Image-to-image translation with conditional adversarial networks," *arXiv Prepr. arXiv1611.07004*, 2016.

[18] G. Antipov, M. Baccouche, and J.-L. Dugelay, "Face Aging With Conditional Generative Adversarial Networks," *arXiv Prepr. arXiv1702.01983*, 2017.

[19] A. Radford, L. Metz, and S. Chintala, "Unsupervised representation learning with deep convolutional generative adversarial networks," *arXiv Prepr. arXiv1511.06434*, 2015.

[20] D. Berthelot, T. Schumm, and L. Metz, "Began: Boundary equilibrium generative adversarial networks," *arXiv Prepr. arXiv1703.10717*, 2017.

[21] D.-A. Clevert, T. Unterthiner, and S. Hochreiter, "Fast and Accurate Deep Network Learning by Exponential Linear Units (ELUs)," *CoRR*, vol. abs/1511.0, 2015.

[22] Z. Liu, P. Luo, X. Wang, and X. Tang, "Deep Learning Face Attributes in the Wild," in *2015 IEEE International Conference on Computer Vision (ICCV)*, 2015, pp. 3730–3738.

[23] OpenCV, "Face Detection using Haar Cascades." .

[24] A. Asthana, S. Zafeiriou, S. Cheng, and M. Pantic, "Incremental Face Alignment in the Wild," in *2014 IEEE Conference on Computer Vision and Pattern Recognition*, 2014, pp. 1859–1866.

[25] X. Cao, Y. Wei, F. Wen, and J. Sun, "Face alignment by Explicit Shape Regression," in *2012 IEEE Conference on Computer Vision and Pattern Recognition*, 2012, pp. 2887–2894.

[26] X. Xiong and F. D. la Torre, "Supervised Descent Method and Its Applications to Face Alignment," in *2013 IEEE Conference on Computer Vision and Pattern Recognition*, 2013, pp. 532–539.

[27] C. Sagonas, G. Tzimiropoulos, S. Zafeiriou, and M. Pantic, "300 Faces in-the-Wild Challenge: The First Facial Landmark Localization Challenge," in *2013 IEEE International Conference on Computer Vision Workshops*, 2013, pp. 397–403.